\documentstyle[12pt]{article}
\def\lb       {\left( }
\def\rb       {\right) }
\def\lmb      {\left\{ }
\def\rmb      {\right\} }
\def\lbb     {\left[ }
\def\rbb      {\right] }
\def\comma      { \, , }
\def\period     { \, . }
\def\bra#1      { \langle \, #1 \, \vert \, }
\def\ket#1      { \, \vert \, #1 \, \rangle \, }
\def\semiket#1  { \, #1 \, \rangle \, }
\def\del        {  \partial  }

\def\Tr      {  \mbox{Tr}  }
\def\abs#1      {  \, \vert #1 \vert \,   }

\def\vecii#1#2      {  \left(\begin{array}{c}#1\\#2\end{array}\right)  }
\def\veciii#1#2#3   {  \left(\begin{array}{c}#1\\#2\\#3\end{array}\right)  }
\def\matrixii#1#2#3#4            {  \left(\begin{array}{cc}#1&#2\\#3&#4
                                       \end{array}\right) }
\def\matrixiii#1#2#3#4#5#6#7#8#9 {  \left(\begin{array}{ccc}#1&#2&#3\\
                                     #4&#5&#6\\#7&#8&#9\end{array}\right)  }
\def\eqabegin         {  \begin{eqnarray}  }
\def\eqaend           {  \end{eqnarray}  }
\def\nn               {  \nonumber  }

\def\parbigskip        {  \par\bigskip  }

\def\sectionnumbering { \setcounter{equation}{0}
         \renewcommand{\theequation}{\arabic{section}.\arabic{equation}}}
\def\appendixnumbering { \setcounter{equation}{0}
         \renewcommand{\theequation}{\Alph{section}.\arabic{equation}}}
\def\mysection#1{\addtocounter{section}{1} \setcounter{subsection}{0}
                 \sectionnumbering 
    {\large\bf \par \bigskip \parbigskip \noindent \arabic{section} \quad  
     #1  }   \par \bigskip \noindent}

\def\mysubsection#1{\addtocounter{subsection}{1}
      \par \bigskip \noindent  {\normalsize\bf
      \arabic{section}.\arabic{subsection} \quad #1  } 
   \par \medskip \noindent }
\def\appsubsection#1{\addtocounter{subsection}{1}
      \par \bigskip \noindent  {\normalsize\bf
      \Alph{section}.\arabic{subsection} \quad #1  } 
   \par \medskip \noindent }
\def\csectionast#1    { \begin{center}  
  \noindent  {\large\bf #1  }   \par \noindent \end{center} }
\def\xxx#1 {{\tt hep-th/#1}}
\def\grqc#1 {{\tt gr-qc/#1}}
\def\npb#1(#2)#3 { Nucl. Phys. {\bf B#1} (#2) #3 }
\def\rep#1(#2)#3 { Phys. Rept.{\bf #1} (#2) #3 }
\def\plb#1(#2)#3{Phys. Lett. {\bf #1B} (#2) #3}
\def\prl#1(#2)#3{Phys. Rev. Lett. {\bf #1} (#2) #3}
\def\prd#1(#2)#3{Phys. Rev. {\bf D#1} (#2) #3}
\def\ap#1(#2)#3{Ann. Phys. {\bf #1} (#2) #3}
\def\rmp#1(#2)#3{Rev. Mod. Phys. {\bf #1} (#2) #3}
\def\cmp#1(#2)#3{Comm. Math. Phys. {\bf #1} (#2) #3}
\def\mpl#1(#2)#3{Mod. Phys. Lett. {\bf A#1} (#2) #3}
\def\ijmp#1(#2)#3{Int. J. Mod. Phys. {\bf A#1} (#2) #3}
\def\mpla#1(#2)#3{Mod. Phys. Lett. {\bf A#1} (#2) #3}
\def\jhep#1(#2)#3{J. High Energy Phys. {\bf  #1} (#2) #3}
\def\cqg#1(#2)#3{Class. Quant. Grav {\bf  #1} (#2) #3}
\def\kket#1   { \vert #1 \rangle \! \rangle }
\def\bbra#1   { \langle \! \! \langle  #1 \vert }
\begin{document}
%
\def\papertitlepage{\baselineskip 3.5ex \thispagestyle{empty}}
\def\preprinumber#1#2#3{\hfill \begin{minipage}{4.2cm}  #1
              \par\noindent #2
              \par\noindent #3
             \end{minipage}}
\renewcommand{\thefootnote}{\fnsymbol{footnote}}
%
%
\papertitlepage
\setcounter{page}{0}
\preprinumber{October 1998}{PUPT-1816}{hep-th/9810135}
\baselineskip 0.8cm 
\vspace{2.0cm}
\begin{center}
{\large\bf BTZ black holes and the near-horizon geometry 
\\ of higher-dimensional black holes}
\end{center}
\vskip 7ex
\baselineskip 1.0cm
\begin{center}
     {\sc Yuji ~Satoh
   \footnote[2]{ysatoh@viper.princeton.edu} }\\
    {\sl Joseph Henry Laboratories, Princeton University \\
 \vskip -1ex
   Princeton, NJ 08544, USA}
\end{center}
\vskip 10ex
\begin{center} {\large\bf Abstract}
\par \vskip 5ex
\parbox{15cm}{
\baselineskip=3.5ex
We investigate the connection between the BTZ black holes and 
the near-horizon geometry of higher-dimensional black holes. 
Under mild conditions, we show that (i) if a black hole has a global 
structure of the type of the non-extremal Reissner-Nordstrom black holes, 
its near-horizon 
geometry is $AdS_2$ times a sphere,  and further 
(ii) if such a black hole is obtained from a boosted black string 
by dimensional reduction, the near-horizon geometry of the latter contains 
a BTZ black hole. Because of these facts, 
the calculation of the Bekenstein-Hawking entropy and the 
absorption cross-sections of scalar fields 
is essentially reduced to the corresponding calculation 
in the BTZ geometry under appropriate conditions. 
This holds even if the geometry is not supersymmetric in the extremal limit. 
Several examples are discussed.
We also discuss some generalizations to geometries which do not have $AdS$ 
near the horizon. 
\par 
%
}
\end{center} 
\newpage
\renewcommand{\thefootnote}{\arabic{footnote}}
\setcounter{footnote}{0}
\setcounter{section}{0}
\baselineskip = 0.6cm
\pagestyle{plain}
\mysection{Introduction}
The three dimensional black holes discovered by Ba{\~ n}ados, Teitelboim
and Zanelli (BTZ) have a very simple geometry which is locally three 
dimensional
anti-de Sitter space ($AdS_3$) \cite{BTZ}. In spite of this, it is known that 
they possess many characteristics of higher-dimensional black holes 
and serve as a useful toy model for black hole physics (for a 
review, see \cite{Carlip}). Recent developments in string theory 
have revealed that they are  more important
than just as a simplified model: Some class of five- and four-dimensional
black holes contains the BTZ black holes in the near-horizon region 
when they are lifted to the black strings in higher dimensions \cite{Hyun,SS}.
Using this fact, it has been shown \cite{SS}-\cite{BL} 
that the microscopic derivation of 
the Bekenstein-Hawking entropy is reduced to 
the corresponding derivation for the BTZ black holes, which has been
discussed in \cite{Carlip2,Strominger,BSS}.\footnote{
For the unsolved problems in \cite{Carlip2,Strominger,BSS}, 
see \cite{Carlip3}.
}   
Such an argument has been extended to other black hole and black string 
geometries in \cite{Teo}-\cite{IZ2}.
Furthermore, by carefully looking at the calculation of the
greybody factors for the $D=5$ and $4$ black holes \cite{MS} 
and by taking into account their connection to the BTZ black holes, 
one realizes that the greybody factors 
effectively come from the near-horizon BTZ geometry.
(For details, see the discussion in section 4). 
Some greybody factors in these cases have also been derived 
\cite{Teo2,KOZ} using the near-horizon
BTZ ($AdS_3$) geometry and the AdS/CFT correspondence 
\cite{adscft}. The AdS/CFT correspondence is
particularly interesting because it might provide the 
fundamental quantum theory of the black holes via the 
near-horizon BTZ black holes.

In addition to these advances, for a large class of non-extremal
black hole and $p$-brane geometries  it has been shown in \cite{YS} that
the scalar wave equations possess features closely 
related to $AdS_3 (SL(2,R))$ and  
that the greybody factors take the
form expected of a CFT in some parameter region. These suggest
that the connections among black holes, the BTZ($AdS_3$)
geometry and some CFT might be extended to more general 
black objects than the above $D=5$ and $4$ black holes.

In this paper, we investigate the connection between the BTZ
black holes and the near-horizon geometry of higher-dimensional
black holes and $p$-branes. In section 2, under mild conditions 
we show that  
(i) if a black hole has a global structure
of the type of the non-extremal Reissner-Nordstrom black holes, 
its near-horizon
geometry becomes $AdS_2$ times a sphere, 
and (ii) if such a black hole is obtained
from a boosted black string by dimensional reduction, 
the near-horizon geometry of the latter contains a BTZ black hole. 
In section 3,
we show that the entropy counting of these 
black holes is reduced to that of the corresponding BTZ black holes
under appropriate conditions. In section 4, we discuss the greybody
factors for scalar fields. We show that (i) 
the scalar wave equations are identical to the $s$-wave equation 
on the BTZ geometry (this holds in all the cases discussed in \cite{YS}),
and further (ii) the greybody factors essentially come from 
the near-horizon BTZ geometry and take the form expected of a CFT
in some parameter region. 
In section 5, we give several examples. They include
the well-known $D=5$ and $4$ black holes with three and 
four charges respectively, the four-dimensional $N=2$ black holes and
dyonic dilaton black holes in generic dimension.
These last dyonic black holes include
the Reissner-Nordstrom black holes themselves as special cases. 
In section 6, we generalize the results in the preceding sections to 
geometries which do not contain $AdS$ near the horizon.
In section 7, we conclude with a summary and a discussion of 
the future directions. The appendix contains
some technical details about the near-horizon geometry of $p$-branes (A.1), 
$AdS_{2}$, $AdS_{3}$ and the BTZ geometry (A.2) and scalar wave equations
and dimensional reduction (A.3).

\mysection{Geometries with two horizons}
In this section, we consider geometries with a global structure of 
the type of the non-extremal Reissner-Nordstrom black holes.
\mysubsection{near-horizon geometries of $p$-brane solutions}
As a preliminary, we study a geometry with 
the metric of the form
\eqabegin
  ds^2_p &=& \biggl( \frac{r}{q} \biggr)^{\alpha d} 
   \lbb - f(r) dt^2 + dy^i dy^i \rbb
    + \lb \frac{q}{r} \rb^{2\beta} f^{-1}(r) dr^2 \comma \nn \\
  f(r) &=& 1- \frac{r_0^d}{r^d}
    \comma \label{nhpbrane}
\eqaend
where $ i = 1, ..., p$.
$p$-brane solutions typically contain this type of geometry near the 
horizon.
$ r_0 \to  0 $ corresponds to their extremal limit.\footnote{
The near-horizon geometries of some class of extremal $p$-branes
have been discussed in detail in \cite{GT,GHT}.
}
We are interested in the case
in which this becomes (locally) $AdS$. 
Since a Lorentzian space becomes locally $AdS$ if and only if  
it has negative constant curvature, (\ref{nhpbrane}) is locally $AdS$ 
only when the tensor 
\eqabegin
  K_{abcd} &=& R_{abcd} -k (g_{ac} g_{bd} - g_{ad} g_{bc})
  \label{Kabcd}
\eqaend
vanishes for a negative constant $k$.
It is straightforward to calculate this for (\ref{nhpbrane}). We display
the result in the appendix. We then find that for $ r_0 = 0$ the geometry
is $AdS_{p+2} $ if $ \beta=1$ ($\alpha,d$ are arbitrary).
Setting $ (r/q)^{\alpha d} = (\omega/c)^2 $ with 
$ c $ a constant, the metric is expressed by Poincar{\` e}
coordinates like (\ref{poincare}). However, for $ r_0 \neq 0$ 
it cannot be $AdS$ in generic dimension. The exceptional cases are $p=0$ and 
$1$: For $p=0$ we have $AdS_2$ when $(\alpha,\beta) = (1,1),(2,1)$.
By a coordinate transformation $ (r/r_0)^d =  z $, 
the metric takes the form of the type (\ref{ads2u}) for $\alpha = 1$
and (\ref{ads2z}) for $\alpha = 2$. 
For $p=1$ we have $AdS_3$ only when $(\alpha,\beta)=(1,1)$. 
By a transformation $ (r/r_0)^d = (\rho/l)^2$, 
the metric is brought to  the form 
of the type (\ref{btz}).

The $AdS_3$ geometry of the type (\ref{poincare}) typically appears 
for extremal black holes and $p$-branes whereas that of the type (\ref{btz})
in non-extremal cases. 
The solutions to the scalar wave equation in each coordinate system
have recently been studied in detail \cite{BKL,KV} in the context of the 
AdS/CFT correspondence \cite{adscft}. It has also been observed that 
the transition from (\ref{poincare}) to (\ref{btz})
causes the thermalization of the associate quantum states \cite{MS2,KV,KOZ}.
Mathematically, these two coordinate systems are naturally related
to the representations of $SL(2,R)$ in the parabolic and hyperbolic basis, 
respectively. The scalar wave equation is nothing but the Laplace 
equation on $SL(2,R)$ and its solutions correspond to the matrix elements
of the representations. They have been well-classified in terms of 
the representation theory. The transition from (\ref{poincare}) to (\ref{btz})
corresponds to a change of the basis and the transition functions (Bogoliubov
coefficients) are also well-studied. For a review, see, e.g., \cite{VK,NS}

\mysubsection{near-horizon geometry of a non-extremal Reissner-Nordstrom type 
\\ \hspace*{2.4em} black hole}
In the following, 
we consider a spherically symmetric geometry in $D(=d+3)$ dimensions,
\eqabegin
   ds^2 &=& - F_1(r) f(r) dt^2 
    + F_2(r) \lbb f^{-1}(r) dr^2 + r^2 d\Omega_{d+1}^2 \rbb
  \comma \label{RNtype}
\eqaend
where 
$d\Omega^2_{d+1}$ is the metric of the unit $(d+1)$-sphere.
If $F_{i}$ $(i=1,2)$ are arbitrary, 
this is the most general form of a spherically 
symmetric geometry. Here we focus on the case in which (\ref{RNtype}) 
expresses
a non-extremal Reissner-Nordstrom type  black hole, namely, in which 
it has two horizons whose $(t,r)$-part  tends 
to become Rindler space near the horizon (up to an overall sign),
\eqabegin 
  && ds^2_{\rm Rindler} = - \kappa^2 \rho^2 dt^2 + d\rho^2
  \comma \label{Rindler}
\eqaend
with $\kappa$ a constant. 
We further concentrate on the case where the outer and inner horizon 
are located at $r=r_0$ and $r=0$ respectively. 
Then $F_i(r_0) $ should be non-vanishing so that the geometry 
contains Rindler space as $r \to r_0$. By a coordinate
transformation $ \rho \sim  \sqrt{r_0(r-r_0)} $, we confirm that
$(t,r)$-part becomes of the form (\ref{Rindler}). 
Similarly, the requirement for $r \to 0$ gives several conditions.
Firstly, since the radius of $S^{d+1}$ should be regular, we have
\eqabegin
    F_2(r) & \to & \lb \frac{r_2}{r} \rb^{2} \qquad \mbox{ as  } \ r \to 0
   \comma 
\eqaend  
with $r_2$ some constant. In addition, $F_1$ should satisfy
\eqabegin
    F_1(r) & \to & \lb \frac{r}{r_1} \rb^{2d} \qquad \mbox{ as  } \ r \to 0
   \period
\eqaend  
Under these conditions, the geometry actually 
takes the form (\ref{Rindler}) times $S^{d+1}$ as $r \to 0 $ 
by the coordinate transformation $ \rho \sim r_0 (r/r_0)^{d/2}$.
For later convenience, we rescale $t$ so that 
$ r_1 = r_2 = q$ and 
rewrite $F_i(r)$ by $F_1 = H_1^{-2}$ and $F_2 = H^{2/d}_2$.
Consequently, the metric is expressed as  
\eqabegin
  ds^2 &=& - H_1^{-2}(r) f(r) dt^2 
   + H_2^{2/d}(r) \lbb f^{-1}(r) dr^2 + r^2 d\Omega_{d+1}^2 \rbb
    \comma \label{RN}
\eqaend
where 
\eqabegin
   H_i(r_0) & \neq & 0 \comma \qquad   
   H_{1,2}(r) \ \to \ \lb \frac{q}{r} \rb^{d} 
   \qquad \mbox{ as  } \ r \to 0 \period
  \label{Hi} 
\eqaend  

In the `near-horizon' region where $H_i(r)$ are well-approximated
by $(q/r)^d $, the geometry becomes
\eqabegin
   \tilde{ds}^2 & \sim & 
  \biggl( \frac{r}{q} \biggr)^{2d} \lbb 1-(r_0/r)^d \rbb dt^2
    + \lb \frac{q}{r} \rb^{2}  \frac{dr^2}{1-(r_0/r)^d}
    + q^2 d\Omega^2_{d+1}
  \period
\eqaend
This corresponds to the $(\alpha,\beta) = (2,1)$ case for $p=0$ 
in the previous section and thus the near-horizon geometry is 
$AdS_2 \times S^{d+1}$.\footnote{
This holds  even in the extremal limit $r_0 \to 0$.
The arguments in the following are also valid in this limit
if it is not singular.
}
Here we remark that the near-horizon region is not necessarily restricted
to $r \sim 0$, which is obvious from (\ref{Hi}). For example, 
we can take $H_i$ to be a product of harmonic functions,
\eqabegin
    H_1(r) &=& H_2(r) \ = \ \prod_i \biggl( 1+\frac{r_i^d}{r^d} \biggr)^{a_i}
   \comma \qquad  r_i^d \ = \ r_0^d \sinh^2 \delta_i
   \label{harmonic}
\eqaend
with $ \sum a_i = 1$. In this case, the `near-horizon' region is given by
$r_i \gg r$.
\mysubsection{near-horizon geometry of the lifted black string}
If a $D$-dimensional black hole solution is obtained 
by dimensional reduction from a $(D+1)$-dimensional solution, 
the corresponding geometry represents a black string. Here
we are interested in a boosted black string geometry,
\eqabegin
   ds_{D+1}^2  & = & \hat{H}^{-1}_1(r) \lbb -dt^2 + dx^2 
   + (r_0/r)^d (\cosh \sigma dt - \sinh \sigma dx)^2 \rbb \nn \\
  && \qquad \qquad \qquad 
    + \hat{H}^{2/d}_2(r) \lbb f^{-1}(r) dr^2 + r^2 d\Omega_{d+1}^2 \rbb
   \comma \label{bstring} 
\eqaend
with $ \sigma $ the boost angle in $(t,x)$-part. 
By the dimensional reduction with respect to $x$,
(\ref{bstring}) reduces to (\ref{RN}) if
\eqabegin
   && \hat{H}_1 = H_1  \lbb H_{1}  h_\sigma^{-1} \rbb^{d/(d+2)}
  \comma \qquad 
   \hat{H}_2  = H_2  \lbb H_{1} h_\sigma^{-1} \rbb^{d/(d+2)}
   \comma \label{Hhat}
\eqaend
where $ h_\sigma(r) = 1+ (r_\sigma/r)^d $ and  
$r_\sigma^d = r_0^d \sinh^2 \sigma $.

In the `near-horizon' region where $ H_i(r) \sim (q/r)^d $ and 
$ h_\sigma(r) \sim (r_\sigma/r)^d $,\footnote{
When $\hat{H}_i $ becomes independent of $r_\sigma$,
the condition $ h_\sigma \sim r_\sigma^d/r^d $ is not necessary. 
See the examples discussed later.
}
the geometry becomes
\eqabegin
   \widetilde{ds}_{D+1}^2 &\sim& \biggl( \frac{r}{\hat{q}} \biggr)^d 
       \lbb -dt^2 + dx^2 + (r_0/r)^d (\cosh \sigma dt - \sinh \sigma dx)^2 \rbb
     \nn \\
     &&   \qquad \qquad  
       + \biggl( \frac{\hat{q}}{r} \biggr)^{2} \frac{dr^2}{1-(r_0/r)^d} 
       + \hat{q}^2 d\Omega_{d+1}^2
      \comma \label{nhbs}
\eqaend
where 
\eqabegin
   &&  
    \hat{q} = q (q/r_\sigma)^{d/(d+2)} \period
   \label{qQ} 
\eqaend
We find that this corresponds to the $(\alpha,\beta) = (1,1)$ case for $p=1$ 
 in section 2.1 and hence
the near-horizon geometry is locally $AdS_3 \times S^{d+1}$. 
Moreover, when $x$ is periodic,
i.e., $ x \sim x + 2 \pi R$,  
this is nothing but the BTZ geometry \cite{BTZ}. Indeed, 
by the coordinate transformation \cite{Hyun,SS,BL}
\eqabegin
   \frac{\rho^2}{R^2} &=& \biggl( \frac{r}{\hat{q}} \biggr)^d + 
                \biggl( \frac{r_0}{\hat{q}} \biggr)^d \sinh^2 \sigma 
       \comma \nn \\
    \varphi & = & \frac{x}{R} \comma 
   \quad \quad \tau \ = \ \frac{l}{R} t
   \comma \quad \quad l \ = \ \frac{2}{d} \, \hat{q} \comma
   \label{nhtoBTZ}
\eqaend
the geometry (\ref{nhbs}) takes  
the standard form of the BTZ metric
\eqabegin
   ds_{BTZ}^2 &=& -N(\rho) d\tau^2 
   + \rho^2 \biggl( d\varphi - \frac{\rho_+ \rho_-}{l \rho^2} d\tau \biggr)^2 
    + N^{-1} (\rho) d\rho^2 \comma \nn \\
    N(\rho) & = & \frac{(\rho^2-\rho^2_+)(\rho^2-\rho^2_-)}{l^2 \rho^2}     
   \comma \label{BTZmetric}
\eqaend
where
\eqabegin
   \rho_+^2 &=& R^2 (r_0/\hat{q})^d \cosh^2 \sigma \comma \qquad
   \rho_-^2 \ = \ R^2 (r_0/\hat{q})^d \sinh^2 \sigma
   \period 
\eqaend
Therefore if the black hole (\ref{RN}) is obtained from 
a boosted black string, the near-horizon geometry of the 
latter contains a BTZ black hole.   
\mysection{Black hole entropy}
In this section, 
we will show that  under some conditions the Bekenstein-Hawking entropy 
of the Reissner-Nordstrom type black holes (\ref{RN}) is obtained from 
the entropy of the BTZ black holes contained in the 
lifted black string geometries.

To see this, let us first note that the entropy of the 
BTZ black hole (\ref{BTZmetric}) is given by
\eqabegin
   S_{BTZ} &=& \frac{A_3}{4 G_3}  
  \ = \ \frac{\pi R}{2 G_3} 
  \biggl( \frac{r_0}{\hat{q}} \biggr)^{d/2} \cosh \sigma
   \comma  \label{Sbtz} 
\eqaend
where $A_{D'}$ and $G_{D'}$ are the area of the outer 
horizon and the Newton constant 
in $D'$ dimensions respectively. We have used $A_3 = 2 \pi \rho_+$. 
On the other hand, the entropy associated to the geometry (\ref{RN}) is
\eqabegin
   S_{D} &=& \frac{A_D}{4 G_D} 
  \ = \ \frac{\omega_{d+1}}{4 G_D} \lbb r_0 H^{1/d}_2(r_0)\rbb^{d+1}
    \comma \label{Srn}
\eqaend
with $\omega_{d+1}$ the volume of the unit $(d+1)$-sphere.

To compare these two expressions, we need to know the relation between the
Newton constants in different dimensions. In general,
under dimensional reduction from $D'+p$ dimensions to $D'$ dimensions, 
the Newton constants in the effective action change as 
\eqabegin
   \frac{1}{G_{D'+p}} &=& \frac{V_p}{G_{D'}}
   \comma \label{dimredG}
\eqaend
where $V_p$ is the volume of the compactified space. This is valid
when the metric is independent of the compactified coordinates. 
Thus we get $ G_{D+1} = G_{D}/2\pi R$ when we lift the black hole to the
black string. Furthermore, since the black string geometry factorizes 
into  the BTZ geometry times $S^{d+1}$ in the near-horizon region, 
the Ricci scalar in the Lagrangian also splits into two parts. 
Hence we obtain a similar relation  to (\ref{dimredG}); 
$G_{D+1} = G_3/(\hat{q}^{d+1} \omega_{d+1}) $. Combining these,
one finds that 
\eqabegin
    \frac{1}{G_D} &=& \frac{2 \pi R}{\hat{q}^{d+1} \omega_{d+1} G_3}
  \period
\eqaend
Substituting this into (\ref{Srn}) yields
\eqabegin
   S_{D} &=&  \gamma S_{BTZ} \nn \\
  \gamma &=& \biggl( \frac{r_0}{\hat{q}} \biggr)^{1+d/2} 
   H_2^{(d+1)/d}(r_0) \cosh^{-1} \sigma
   \period
\eqaend
Thus the two entropies agree when 
\eqabegin
   && \gamma \sim 1 \period \label{gamma1}
\eqaend
This conditions is typically satisfied  
when the radii of the outer and inner horizon are close and 
$r_\sigma \gg r_0$ $(\sigma \gg 1)$.
In fact, this means that 
$  r_0 H_2^{1/d}(r_0) \sim \lim_{r \to 0} r H_2^{1/d}(r) \sim q$. 
Using (\ref{qQ}), we find that
\eqabegin
 && \hat{q} \sim r_0 
    \lbb H_2^{2(d+1)/d}(r_0) \sinh^{-2} \sigma \rbb^{1/(d+2)}  
   \comma
\eqaend 
and (\ref{gamma1}) is satisfied.
In addition, it may be useful to evaluate $\gamma$ for the example given in 
(\ref{harmonic}). It is given by
\eqabegin
   \gamma &=& 
  \prod_i \lb \frac{\cosh \delta_i}{\sinh \delta_i} \rb^{2a_i \frac{d+1}{d}}
    \frac{\sinh \sigma}{\cosh \sigma}
   \period \label{eta}
\eqaend 
Thus $S_{D}$ is obtained from $S_{BTZ}$, e.g., when 
$ \delta_i, \sigma \gg 1$, 
namely, the charges $r_i, r_\sigma $ are large enough or the 
geometry is near the 
extremal limit $r_0 \to 0; \delta_i, \sigma \to \infty$.
We remark that when
\eqabegin
    && 2 a_j (d+1) = d \comma \qquad \sigma = \delta_j \comma
   \label{sigmaindep} 
\eqaend
for some $j$, $\hat{H}_i$ and $\hat{q}$ become independent of 
$ \sigma = \delta_j$ and we do not need the condition for $r_\sigma$.

For the BTZ black holes, the microscopic derivation 
of the Bekenstein-Hawking
entropy has been discussed in \cite{Carlip2,Strominger,BSS}.
Our argument here indicates  that the microscopic derivation of 
the entropy of the Reissner-Nordstrom type black holes in (\ref{RN})
is reduced to the corresponding derivation 
in \cite{Carlip2,Strominger,BSS} under appropriate conditions.
\mysection{Greybody factors}
We have discussed the Bekenstein-Hawking entropy.
In this section, we discuss another important feature of
black holes, i.e., greybody factors. We will find that they
essentially come from the near-horizon BTZ geometry under mild conditions.

We begin with a brief review of the results of \cite{YS}. 
In \cite{YS}, the propagation
of minimally coupled massless scalars in various non-extremal 
black hole and $p$-brane geometries has been
 discussed. There it has been  shown that by expanding a scalar field as 
$ \Phi = \sum \ e^{-i\omega t} \phi_m(r) \chi_m$, 
the near-horizon equation for the $m$-th partial wave takes the form 
\eqabegin
  \lbb \Bigl\{ z(z-1)\del_z \Bigr\}^2 - z(z-1) 
  \biggl\{ \frac{\nu_+^2}{z-1} - 
  \frac{\nu_-^2}{z}  + \frac{\Lambda}{d^2} \biggr\} \rbb \phi_m &=& 0
  \period \label{hyper}
\eqaend
Here, 
$\chi_m$ is the eigenfunction of the Laplacian on $S^{d+1}$ 
with the eigenvalue $-\Lambda = -m(m+d)$;
$\nu_\pm = -i \beta_\pm \omega/4\pi $; $ \omega $ is the frequency; 
$\beta_\pm $ are the periods of the imaginary time associated to the outer 
and inner horizon (when the geometry does not have an inner horizon, 
$\beta_- = 0$). In addition, the above equation has been identified with the 
scalar wave equation in $ SL(2,R) $ ($AdS_3$). 
One finds this as follows.  
The eigenvalue equation of the Laplace operator on $SL(2,R)$ 
is given by 
\eqabegin
   \lbb \frac{1}{\sqrt{-g}} \del_\mu \sqrt{-g} g^{\mu \nu} \del_\nu 
   -4j(j+1) l^{-2} \rbb \ \Phi &=& 0 
  \comma \label{sl2}
\eqaend
where $j$ is the $SL(2,R)$ spin and 
$l$ is a parameter in terms of which the scalar curvature is
written as $-6l^{-2}$. Here we use the coordinate system in (\ref{btz})
and make a change of variables $ z= (r/l)^2 $. As $z \to 1$ and $0$, 
the geometry approaches Rindler space. We further rescale $ \theta_\pm$
as $ \tilde{\theta}_\pm = (\tilde{\beta}_\pm/2\pi l) \theta_\pm $ 
so that their 
imaginary periods associated to the Rindler space-like regions 
become $ \beta_\pm $. By a separation of variables
$ \Phi = \sum \exp(i\omega_+ \tilde{\theta}_+ + i \omega_- \tilde{\theta}_-) 
\phi(r) $, the radial equation for $\phi$ takes the same form 
as (\ref{hyper}) under the identifications
\eqabegin
   \beta_\pm \ \leftrightarrow \ \tilde{\beta}_\pm \comma \qquad 
   -(1+m/d) \ \leftrightarrow \ j \comma \qquad 
   \abs{ \omega }  \ \leftrightarrow \ \abs{ \omega_\pm }
   \period \label{ident}
\eqaend
This means that the near-horizon wave equations  
for the various black objects
are regarded as those in the $SL(2,R)$ background.  

Now we show that, more precisely, (\ref{hyper}) is naturally identified with 
the $s$-wave equation on the BTZ background. The point is the last 
identification in (\ref{ident}). For definiteness, we fix the 
signs and take it to be $ \omega = \omega_\pm $. This indicates that 
we are dealing with the wave function 
$ \Phi 
= \sum \phi(r) \exp \Bigl[ i \omega (\tilde{\theta}_+ + \tilde{\theta}_-) 
\Bigr]$.
Here let us recall that the time and angle coordinate of the BTZ metric
(\ref{BTZmetric}) are given by (\ref{tauphi}). In terms of the periods of 
the imaginary time associated to the two horizons,
\eqabegin
   \tilde{\beta}_\pm &=& \frac{2 \pi l^2 \rho_\pm}{\rho_+^2 - \rho_-^2}
   \comma \label{beta}
\eqaend
they are given by
\eqabegin
   \vecii{\tau}{l \varphi} &=& \frac{1}{2\pi l} 
  \matrixii{\tilde{\beta}_+}{\tilde{\beta}_-}{\tilde{\beta}_-}{\tilde{\beta}_+}
   \vecii{\theta_+}{\theta_-}
 \period \label{tautheta}
\eqaend
Therefore, $ \tilde{\theta}_+ + \tilde{\theta}_- = \tau $ and 
$ \Phi = \sum \phi(r) 
 \exp \Bigl[ i \omega (\tilde{\theta}_+ + \tilde{\theta}_-)
  \Bigr] = \sum \phi(r) \exp(i \omega \tau + i n \varphi)\ \vert_{n=0} $.
Namely, the wave equation is the $s$-wave equation on the BTZ background. 

Let us turn to the wave equation of a massless scalar 
in the geometry (\ref{RN}). Using the same separation of variables as 
for (\ref{hyper}), one obtains the radial equation,  
\eqabegin
  \lbb \lmb (H_2/H_1) f r^{d+1} \del_r \rmb^2 
   + H_2^{2(d+1)/d} r^{2(d+1)} \omega^2
   - (H_2/H_1)^2 f r^{2d} \Lambda \rbb \phi_m(r) &=& 0
  \period \label{Eqrn}
\eqaend
In the near-horizon region where $H_1 \sim H_2 \sim (q/r)^d $,
a change of variables $(r/r_0)^d = z $ yields  
the wave equation of the form (\ref{hyper})
with $\nu_+^2 \sim \nu_-^2 = - (q/r_0)^{2(d+1)} (r_0 \omega/d)^2 $.
Thus it is identical to  the $s$-wave equation on the BTZ background
as discussed above.\footnote{
To directly compare the wave equation from the near-horizon BTZ geometry
with (\ref{Eqrn}) near the horizon, in addition to (\ref{ident})
we need to take into account the 
rescale of the time coordinate in (\ref{nhtoBTZ}), i.e., 
the rescale of the frequency $\omega \to (l/R) \omega $.
} 

In section 2, we saw that 
the near-horizon geometry of (\ref{RN}) is not  
$AdS_3 \times S^{d+1}$ but $AdS_2 \times S^{d+1}$. 
Then why does the wave equation in the geometry 
(\ref{RN}) become the $s$-wave equation in the BTZ background ? 
One can understand this as follows.
When the black hole is lifted
to the black string, the near horizon geometry of the latter 
contains the BTZ geometry.
In addition, the radial wave equation is invariant under the dimensional
reduction if the wave function is independent of the reduced coordinate
(see, the appendix). Thus the near-horizon equation for (\ref{RN}) is 
the same as the corresponding equation for the black string 
and hence it becomes the wave equation
in the BTZ geometry. Moreover, since the reduced coordinate is nothing but
the angle coordinate of (\ref{BTZmetric}), being independent of the 
reduced coordinate means taking the $s$-waves in terms of the BTZ geometry.

Having established the relation between (\ref{Eqrn}) and (\ref{hyper}), 
 similarly to \cite{MS3,YS} one can show that
the greybody factors take the form 
expected of a CFT at low energy.  
To see this, we need only mild assumptions such as asymptotic
flatness ($H_i \to 1$ as $r \to \infty$), $ q/r_0 \gg 1$ (`large charge')
and so on. In particular, when the geometry is given by (\ref{harmonic}),
the argument is almost the same as in \cite{YS}.\footnote{
According to the form of $H_i$, more precise approximation is of course
possible as extensively studied in the literature, e.g., \cite{MS}. 
} 

Furthermore, we realize that the greybody factors essentially come form 
the near-horizon BTZ geometry. This is because
(i) the essence of the calculation of the absorption cross-sections 
is the matching 
between the wave functions in the near-horizon and the asymptotically 
flat region, (ii) the non-trivial energy-dependence (greybody factors)
comes from the former and (iii) near the horizon 
the wave equation (\ref{Eqrn}) becomes the equation in the  
BTZ geometry contained in the lifted black string geometry. 
Thus the calculation of the greybody factors for the geometry 
(\ref{RN}) becomes essentially the same as the corresponding
calculation for the BTZ geometry discussed in \cite{BSS2}.
It may be worth noting that the greybody factors obtained 
in \cite{BSS2} take the form expected of a CFT only 
for the $s$-waves. 

Therefore, we see that the BTZ black holes play an important role 
in the study of Hawking radiation for the Reissner-Nordstrom type 
geometry (\ref{RN}). 
Taking into account the result in appendix A.3, 
we also see that the argument in this section can be generalized to 
the charged scalar case. 

It is subtle to explain the BTZ structure in (\ref{Eqrn})
directly in terms of the near-horizon geometries.
When we dimensionally reduce (\ref{BTZmetric}) (in string frame)  
with respect to $\varphi$,  we obtain the metric 
\eqabegin
  ds_2^2 &=& -N(\rho) dt^2 + N^{-1}(\rho) d\rho^2
  \comma \label{NN}
\eqaend 
with the dilaton $ e^{-2\Psi} = \rho $. We cannot absorb the dilaton and go
to Einstein frame since we are in two dimensions.
By a coordinate transformation $z= (\rho^2 - \rho^2_-)/\Delta^2$ with 
$\Delta^2 = \rho^2_+ - \rho^2_-$, the metric becomes
\eqabegin
   ds^2_2 &=& 
  - \lb \frac{\Delta}{l} \rb^2 \frac{z(z-1)}{z + (\rho_-/\Delta)^2 }
     dt^2 + \frac{l^2}{4} \frac{dz^2}{z(z-1)}
   \period
\eqaend
This is not $AdS_2$ contrary to the near-horizon geometry of (\ref{RN}). 
It approaches 
$AdS_2$ when $ z \ll \rho^2_-/\Delta^2$. This corresponds to 
$ \rho \to \rho_- $ or $ \rho_- \to \rho_+ $ with $ z $ fixed.
In the latter case, it follows from (\ref{tautheta}) that 
$ \varphi \propto \theta_L$ where 
$ \theta_L = (\theta_+ + \theta_-)/l $ is an analog of an Euler angle
associated to the left $SL(2,R)$.  Then the above dimensional 
reduction becomes that discussed in \cite{LS,Strominger2} which 
relates $AdS_3$ to $AdS_2$ (see also the appendix). The argument here
is essentially the same as in \cite{Strominger2} in the extremal case. 
Though the scalar wave equation in (\ref{NN}) does not take the form
(\ref{hyper}),  the tachyon wave 
equation $ \Bigl[(\sqrt{-g} e^{-2\Psi})^{-1} 
\del_\mu \sqrt{-g} e^{-2\Psi} g^{\mu \nu} \del_\nu - \mu^2 \Bigr] \Phi= 0 $
does.
%
%
\mysection{Examples}
We argued that the entropy and the greybody factors of  
the Reissner-Nordstrom type black holes (\ref{RN}) effectively come from the
near-horizon BTZ black holes. In this section, we give several examples.

The first examples of the geometry of the type
(\ref{RN}) are the $D=5$ and $4$ black holes in string theory.
The metrics are given respectively by (see, e.g., \cite{Youm})
\eqabegin
   && H^{(5)}_1 = H^{(5)}_2 = \prod_{i=1}^3 
    \biggl( 1+ \frac{r^2_i}{r^2} \biggr)^{1/3} 
  \comma \qquad 
    H^{(4)}_1 = H^{(4)}_2 
       = \prod_{i=1}^4 \biggl( 1+ \frac{r_i}{r} \biggr)^{1/4} \comma
   \label{54bh} 
\eqaend
where $ r_i^d = r_0^d \sinh^2 \delta_i$. 
To get the corresponding black string geometry, we set, e.g., $r_1 = r_\sigma$
in (\ref{Hhat}) so that 
\eqabegin
   \hat{H}^{(5)}_{1,2} 
     = \prod_{i=2}^{3}  
    \biggl( 1+ \frac{r^2_i}{r^2} \biggr)^{1/2} \comma \qquad
   \hat{H}^{(4)}_{1,2}  
     = \prod_{i=2}^{4}  \biggl( 1+ \frac{r_i}{r} \biggr)^{1/3}  
  \period \label{H45hat}
\eqaend 
Note that $r_\sigma$-dependence has disappeared.

In this case, the microscopic derivation of the Bekenstein-Hawking 
entropy has been discussed in \cite{SS}-\cite{BL} using the 
near-horizon BTZ geometry. 
Since the condition (\ref{sigmaindep}) holds, 
the condition $ \gamma \sim 1$ in section 3 is satisfied  
irrelevantly to $r_\sigma$ if $ \delta_{j} \gg 1 $ 
($j=2,3$ for $D=5$ and $j =2,3,4$ for $D=4$).
Note that this includes the dilute gas region $ r_0, r_1 \ll r_j$.

As for the greybody factors, the fact that they essentially come from 
the near-horizon BTZ geometry has already been implied in the literature:
we have only to repeat the discussion in the previous section.
This was explicitly confirmed in the  $D=5$ case 
by replacing the near-horizon geometry with the BTZ geometry
\cite{LM}. We also find that the $ SL(2,R)$ structure of the
wave equations observed in \cite{CL} is due to the near-horizon
BTZ geometry.
The greybody factors in these cases have been 
derived using the AdS/CFT correspondence \cite{Teo2,KOZ}.

The next example is a class of the black holes in 
the four-dimensional $N=2$ supergravity \cite{n=2}.
The geometry is given by
\eqabegin
   && H^{N=2}_1 = H^{N=2}_2 =  \bigl( h_g d^{ABC} h_A h_B h_C
   \bigr)^{1/4} 
   \comma
\eqaend
where 
$A,B,C = 1, ..., n_v$; $ h_{g,A}(r) = b_{g,A}(1+ r_{g,A}/r) $;
$ r_{g,A} = r_0 \sinh^2 \delta_{0,A}$; $b_{g,A}$ are constants
satisfying $b_g d^{ABC} b_A b_B b_C = 1$; 
$n_v$ is the number of the vector multiplets
(not including the universal one); 
and $d^{ABC}$ are the topological intersection numbers of the Calabi-Yau
manifold.
The corresponding black string solution is obtained by setting 
$ r_\sigma = r_g $ \cite{IZ2};
\eqabegin
   && \hat{H}^{N=2}_{1,2} = 
   \bigl( b_g d^{ABC} h_A h_B h_C \bigr)^{1/3}
   \period  
\eqaend
In this case, the microscopic entropy counting using the near-horizon BTZ
geometry has been discussed 
in \cite{IZ2}. Similarly to \cite{MS,MS3,YS}, one can discuss the 
greybody factors. The entropy counting for five-dimensional
$ N=2$ extremal black holes has also been discussed in \cite{IZ2}.

Now we discuss new examples.
Here we consider a class of dyonic dilaton black holes
\cite{GM}. The metric is given by (\ref{RN}) with\footnote{
We use slightly different notations from those in \cite{YS};
the relation is $ r^d \leftrightarrow r^d + \eta_0$, 
$ r_0^d \leftrightarrow 2\eta_0 $ and so on.
}
\eqabegin
   H^{\rm Dy}_{1} = H^{\rm Dy}_2  
   = \lbb \biggl( 1+ \frac{r^d_1}{r^d} \biggr)^{\alpha_1}
          \biggl( 1+ \frac{r^d_2}{r^d} \biggr)^{\alpha_2} \rbb^{1/2} 
  \comma \label{dybh}
\eqaend 
where $\alpha_i$ are some parameters satisfying $ \alpha_1 + \alpha_2 = 2$.
This includes the Reissner-Nordstrom black holes 
in generic dimension as the special case $r_1 = r_2$. 
Note that generically they are not supersymmetric
even in the extremal limit $r_0 \to 0$. We further focus on the 
case of $ \alpha_1 = d/(d+1) \comma \alpha_2 = (d+2)/(d+1)$.\footnote{
Most of the discussion in the previous sections  formally
holds for generic $\alpha_i$. However, the physical interpretation is unclear
unless we explicitly construct the lifted black string solution.
}
$d=1$ and $2$ give special cases of  (\ref{54bh}).
By setting $r_\sigma = r_1$ in (\ref{Hhat}), we obtain the lifted black string
geometry, 
\eqabegin
  ds^2_{bs} &=& 
    \biggl( 1 + \frac{r_2^d}{r^d} \biggr)^{-1} \lbb -dt^2 + dx^2 
   + (r_0/r)^d (\cosh \sigma dt - \sinh \sigma dx)^2 \rbb \nn \\
  && \qquad \qquad \qquad 
    + \biggl( 1 + \frac{r_2^d}{r^d} \biggr)^{2/d}
     \lbb f^{-1}(r) dr^2 + r^2 d\Omega_{d+1}^2 \rbb 
  \comma \label{ndbs}
\eqaend
which corresponds to $
  \hat{H}^{\rm Dy}_{1,2} = 1+ (r_2/r)^d $.
This is nothing but the boosted geometry of the non-dilatonic
black string solution by \cite{GHT}.
The entropy counting for this class of dyonic black holes
is again reduced to that of the BTZ black hole.
Since the condition (\ref{sigmaindep}) again holds,  
$\gamma \sim 1$ is satisfied if $ r_2 \gg 1$.  
The greybody factors for these geometries have been discussed in 
\cite{YS}. 
\mysection{Generalizations to other black holes and $p$-branes}
The argument in the previous sections 
can be generalized to some cases in which the near-horizon geometry is 
not directly related to $AdS$. 

Let us first note that
because of the relation between the Newton constants 
in different dimensions (\ref{dimredG}), the Bekenstein-Hawking entropy
is invariant under dimensional reduction;
\eqabegin
   \frac{A_{D}}{4G_{D}} &=& \frac{A_{D+p}}{4G_{D+p}}
   \period \label{BHentropy}
\eqaend  
In addition, the radial wave equation is also invariant (as discussed 
in the appendix). On the other hand, 
the near-horizon geometry is {\it not };
even if a geometry contains $AdS$ near the horizon, generically 
the near-horizon geometry connected  by dimensional reduction does not contain 
$AdS$.
These allow us to reduce the discussion on the entropy and 
greybody factors to that for the BTZ black holes even for 
the geometries without near-horizon $AdS$.

In the following, we will focus on a class of $p$-branes 
in $D$ dimensions \cite{DLP},\footnote{
This corrects typos in \cite{YS}.
}
\eqabegin
 ds_{pB}^2 &=& H^\alpha (r) \biggl( H^{-N}(r) 
    \lbb -f(r) dt^2 + dy^i dy^i \rbb
        + f^{-1} (r) dr^2 + r^2 d\Omega_{d+1}^2 \biggr)
   \comma \nn \\
  && H(r) \ = \ 1 + \frac{Q}{r^d} \comma 
     \qquad \alpha \ = \ \frac{p+1}{d+p+1} N \comma \label{dlp}
\eqaend
where $ i = 1, \cdots, p$, $D=d+p+3$ and $N$ is some parameter. 
The extremal case ($ r_0 = 0 $) is supersymmetric if and only 
if $N$ is an integer. 
For later use, we introduce another parameter
\eqabegin
  \xi &=& 1 + \frac{1}{d} - \frac{N}{2}
  \period 
\eqaend
The dimensional reduction with respect to $y^i$ connect 
the geometries with different $p$ but $N$ and $\xi$ are invariant. 
The Reissner-Nordstrom black holes discussed in the previous section
are included in (\ref{dlp}) and correspond to $ \xi = 0, p=0 $. 

We first discuss a class of magnetically charged 
black holes by \cite{HS}, which is a special case of
(\ref{dlp}) with $ \xi = 1/2 \comma  p = 0$;
\eqabegin
   ds^2_{\rm mag} &=& 
  - \biggl( 1+\frac{Q}{r^d} \biggr)^{-(d+2)/(d+1)} f(r) dt^2 + 
   \biggl( 1+\frac{Q}{r^d} \biggr)^{(d+2)/d(d+1)} 
   \lbb f^{-1}(r) + r^2 d\Omega^2_{d+1} \rbb
   \period \nn \\
    && \label{magbh}
\eqaend
From the argument in section 2, one finds that the 
near-horizon geometry of (\ref{magbh}) does not contain $AdS$. However, 
(\ref{magbh}) is obtained by dimensional reduction from an un-boosted 
non-dilatonic black string, namely, from 
(\ref{ndbs}) with $\sigma = 0$ and $ r_2^d = Q$. 
(This is also a special case of (\ref{dlp}) 
with $\xi = 1/2, p=1  $).  Then as discussed above, the discussion
on the greybody factors again reduces to that on the 
near-horizon BTZ geometry. 

Although the argument in section 3 becomes singular when $ \sigma = 0$, 
it can be applied to this case with appropriate modification. 
First, it follows from (\ref{BHentropy}) that 
the Bekenstein-Hawking entropy of (\ref{magbh}) is equal to that 
of the black string (\ref{ndbs}) with $\sigma = 0$. Second, 
the entropy of (\ref{ndbs}) with $\sigma = 0$ and the radius of $x$ equal
to $R$ is the same as that of (\ref{ndbs}) with $\sigma \neq 0$ and 
the Lorentz contracted radius $ R' = R/\cosh \sigma $. 
This means that the entropy of the black string, which is supposed to 
express the total number of the relevant states, is invariant
under boosts. For more detailed discussions on the entropy and boosts, see 
\cite{boost}. Since by dimensional
reduction the geometry (\ref{ndbs}) with $\sigma \neq 0$ gives 
the dyonic dilaton black holes (\ref{dybh}), the argument 
in the previous section is applied to the boosted black string. 
To discuss the original entropy of (\ref{magbh}), we 
have only to use $R'$ in the previous argument. 
We remark that one cannot freely boost the black strings in the examples in 
the previous section
since $ \sigma$'s were fixed by the relation $ r_\sigma = r_{1,g}$.  
The condition
(\ref{gamma1}) is satisfied when $ Q \gg 1$.
The entropy counting for $d=1$ case has been discussed in a somewhat 
different way in \cite{IZ}.

Together with the previous argument on the Reissner-Nordstrom black holes, 
we now understand 
the connection between the BTZ black holes and the geometries with 
$(\xi,p) = (0,0)$ and $(1/2,0)$. Furthermore, since the geometries with 
different $p$ are related by dimensional reduction, it is possible to 
similarly discuss the entropy and the greybody factors for 
all the cases with 
\eqabegin
   \xi &=& 0 \quad \mbox{ or } \quad \frac{1}{2}
   \period
\eqaend 

In \cite{YS}, the scalar wave equation for (\ref{dlp}) has been 
analyzed and some features closely related to $AdS_3$ have been 
found. The discussion here explains their geometrical reason for 
$\xi = 0,1/2$. 

Starting from (\ref{bstring}) with (\ref{H45hat}) and $\sigma = 0$, 
dimensional reduction yields the black holes in string theory 
in $D=5$ and $D=4$ with two and three charges respectively \cite{Youm}.
One can discuss the entropy and the greybody factors similarly 
to the case of (\ref{magbh}). In fact, if all these charges are
the same, the geometries become (\ref{magbh}). 
The corresponding greybody factors have been discussed in \cite{YS}. 
The geometry in the $D=5$ case is closely related to the fundamental 
strings and the entropy in the extremal case 
has been discussed using the stretched horizon \cite{PS}. 
\mysection{Discussion}
We argued that under mild conditions 
if a black hole geometry has a global structure of the type of 
the non-extremal Reissner-Nordstrom black holes, 
its near-horizon geometry becomes $AdS_2$ times a sphere.
In addition, if such a black hole can be lifted to a boosted black string,
the near-horizon geometry of the latter contained a BTZ geometry.
Because of these facts, the calculation of the Bekenstein-Hawking
entropy and the greybody factors was essentially reduced to the 
corresponding calculation  in the 
near-horizon BTZ geometry under appropriate conditions. 
Furthermore, using dimensional reduction, 
we showed that in some cases similar arguments hold even if the geometry 
does not contain $AdS$ near the horizon.

The arguments in this paper extend the results in the literature
on the black hole thermodynamics 
using the near-horizon BTZ geometry 
\cite{Hyun}-\cite{BL},\cite{Teo}-\cite{IZ2},\cite{LM}. 
In contrast to the cases discussed in the literature, 
ours include the geometries whose extremal limit is not 
supersymmetric as the Reissner-Nordstrom black holes in $D \neq 4,5$.

We also showed that the near-horizon wave equations discussed in 
\cite{YS} are nothing but the $s$-wave equation in the BTZ background
and gave a geometrical explanation for the $AdS_3$ structure
found in \cite{YS} in some cases. 
By further investigations, the connection to 
the BTZ black holes might be found also in the other cases.  

Our results indicate that there exists a kind of universality
for the Reissner-Nordstrom type black holes (\ref{RN}) as pointed out in 
some cases by \cite{Hyun}; the thermodynamics 
is effectively reduced to that of the BTZ black holes
and its characteristics do not depend on 
details of each black hole. This is analogous to  
the no-hair theorem and the universal behavior of the absorption 
cross-sections found in \cite{DGM,YS}.
It is suggestive that the conformal diagrams for the BTZ black holes 
are similar to those of the Reissner-Nordstrom black holes. 

Moreover, the CFT structure of the greybody factors discussed in 
section 4 and \cite{YS} suggests
existence of the underlying CFT describing the black holes and 
$p$-branes. 
The near-horizon BTZ geometry may give important clues.
In particular, it would be very interesting to further investigate 
the string theory and quantum gravity 
in $AdS_3$ and $AdS_2$ \cite{Strominger2} and 
the connection to the AdS/CFT correspondence \cite{adscft} 
discussed in \cite{Teo2,KOZ}. 

Finally, for the black holes in (\ref{54bh}) in the toroidally 
compactified string theory, 
the correspondence between the black holes and CFT's 
(including the AdS(BTZ)/CFT correspondence) have been extensively studied.
The argument in section 4 implies that 
such a correspondence might be extended
to the $N=2$ black holes in the Calabi-Yau compactification.
In this case, we have the exactly solvable CFT describing the 
string theory, i.e., Gepner models. It may be another interesting problem to 
explore the $N=2$ black holes in relation to the near-horizon
BTZ black holes and the Gepner models.
   
%
%
\newpage
\csectionast{Acknowledgements}
I would like to thank V. Balasubramanian, S.S. Gubser, K. Hori, 
A.W. Peet and D. Waldram for useful discussions and/or conversations.
I would also like to thank the Aspen Center for Physics, 
where part of this work was done, for its hospitality. This work was
supported in part by Japan Society for the Promotion of Science.
%
%
   \setcounter{section}{1}
\appendixnumbering
\par \bigskip \parbigskip 
\noindent
{\large\bf Appendix}
%
\appsubsection{result of $K_{abcd}$}
Using computers, it is easy to calculate the tensor defined in (\ref{Kabcd}).
The non-vanishing independent components for the geometry (\ref{nhpbrane}) are
\eqabegin
  K_{trtr} &=& \frac{1}{4} r^{-(2+d)} (r/q)^{\alpha d } 
    \Bigl[ - d(\alpha-1) \{ d(\alpha-2) +2 (\beta-1) \} \mu \nn \\  
  && \qquad \qquad \qquad \qquad \left. 
     + \alpha d \{ \alpha d + 2(\beta-1) \} r^{d} + 4k q^{2\beta} 
  r^{2(1-\beta)+d} 
   \rbb \comma \nn \\ 
K_{t y_i t y_i} &=& \frac{1}{4} r^{-2(1+d)} (r/q)^{2(\alpha d+\beta)} 
    \lbb \alpha(\alpha-1)d^2 \mu^2 + \alpha(1-2\alpha)d^2 \mu r^d
   \right. \nn \\
  && \qquad \qquad \qquad \qquad \left. 
    - 4 k q^{2 \beta} \mu r^{2(1-\beta) +d} + \alpha^2 d^2 r^{2d} 
  + 4k q^{2\beta} r^{2(1-\beta+d)} 
   \rbb \comma  \\
K_{y_i r y_i r} &=& \frac{1}{4} r^{-2} (r^d-\mu)^{-1} (r/q)^{\alpha d} 
    \Bigl[ \alpha d \{ 2(\beta-1) + d(\alpha-1) \} \mu
     \nn \\
 && \qquad \qquad \qquad \qquad \left.  
    - \alpha d \{ \alpha d + 2(\beta-1) \} r^d - 4kq^{2\beta} r^{2(1-\beta) +d}
   \rbb \comma \nn \\
K_{y_i y_j y_i y_j} &=& \frac{1}{4} r^{-(2+d)} (r/q)^{2(\alpha d+\beta)} 
    \lbb \alpha^2 d^2 (\mu-r^d) - 4kq^{2\beta} r^{2(1-\beta) +d}
   \rbb \qquad (i \neq j) 
  \comma \nn
\eqaend
with $\mu = r_0^d$. 
We have only $K_{trtr}$ for $p=0$ and $K_{trtr}, K_{tyty} $ and $ K_{yryr} $ 
$(y_1 \equiv y) $ for $p=1$.
%
\appsubsection{$AdS_2$, $AdS_3$ and the BTZ geometry}
Here we briefly summarize the geometries of $AdS_2$, $AdS_3$ 
and the BTZ black holes and their relation.

$AdS_{2+p}$ is defined by
\eqabegin
&&   -x_0^2 -x_1^2 + \sum_{i =2}^{p+1} x_i^2 = - l^2 \comma \nn \\
&&  ds^2 = -dx_0^2 - dx_1^2 + \sum_{i = 2}^{p+1} dx_i^2 
  \period
\eqaend 
A part of $AdS_2$ is parametrized by
\eqabegin
   && x_0 = \sqrt{r^2-l^2} \sinh \theta \comma \qquad x_1 = r \comma 
    \qquad x_2 = \sqrt{r^2 - l^2} \cosh \theta 
  \period 
\eqaend
Then the  metric takes the form
\eqabegin
   ds^2 &=& -(r^2 -l^2) d\theta^2 + [(r/l)^2 -1]^{-1} dr^2  \\
        &=& l^2 \biggl[ -(u-1) d\theta^2 + \frac{1}{4} \frac{du^2}{u(u-1)}
                \biggr] 
           \label{ads2u} \\
        &=& l^2 \biggl[ -4z(z-1) d\theta^2 +  \frac{dz^2}{z(z-1)}
           \biggr]  \label{ads2z}
   \period
\eqaend 
Here we have defined $u$ and $z$ by $u= r^2/l^2$ and 
$ u-1 = 4z(z-1)$ respectively.

To parametrize $AdS_3 (SL(2,R))$, 
the following forms of the metric are often used,
\eqabegin
   ds^2 &=& - \cosh^2 \mu \ d\theta_+^2 + d\mu^2 + \sinh^2 \mu \ d\theta_-^2
           \\
        &=& \frac{l^2}{\omega^2} 
         \lb d\omega^2 - dx^+ dx^- \rb \label{poincare} 
      \\
        &=& - \lbb \lb r/l \rb^2 - 1 \rbb d\theta_+^2 
           +  \lbb \lb r/l \rb^2 - 1 \rbb ^{-1} 
             dr^2 + \lb r/l \rb^2 d\theta_-^2
   \period  \label{btz}
\eqaend
In terms of $SL(2,R)$, these correspond respectively to the parametrization
 of $g \in SL(2,R)$, 
\eqabegin
   && g  =   g_0(\theta_L) g_1(2\mu/l) g_0 (\theta_R) \comma 
    \quad  \! \!
       g_3(x^-/l) g_2(\tau) s g_3(x^+/l) \comma \quad \! \!  
       g_2(\theta_L) g_1(\rho) g_2(-\theta_R)
   \comma
\eqaend 
and $ds^2 = (l^2/2) \Tr (g dg^{-1} g dg^{-1})  $, where  
\eqabegin
   && g_0(t) = \matrixii{\cos t/2}{\sin t/2}{-\sin t/2}{\cos t/2} 
        \comma \qquad   
      g_1(t) = \matrixii{\cosh t/2}{\sinh t/2}{\sinh t/2}{\cosh t/2} 
       \comma \nn \\ 
    && g_2(t)  = \matrixii{e^{t/2}}{0}{0}{e^{-t/2}} \comma \qquad   
      g_3 (t) = \matrixii{1}{0}{t}{1} \comma \qquad 
     s = \matrixii{0}{1}{-1}{0} \comma 
\eqaend
$ \theta_\pm = (l/2) (\theta_L \pm \theta_R)$, $\omega = l \ e^{-\tau/2}$ 
and $ r^2 = l^2 \cosh^2 (\rho/2)$.

These parametrizations are closely related to the $SL(2,R)$ representations
in the elliptic, parabolic and hyperbolic basis, respectively \cite{VK}.
The scalar wave functions in each coordinate system are nothing but 
the matrix elements of the representations in the corresponding basis. 
They are well-classified in terms of the representation theory and typically 
expressed by the Jacobi polynomials, the Bessel functions and the 
hypergeometric functions, respectively. 

To get the BTZ black hole geometry, the parametrization (\ref{btz}) is used.
By making the changes of variables
\eqabegin
   \frac{r^2}{l^2} &=& \frac{\rho^2 - \rho^2_-}{\rho_+^2 - \rho_-^2} \comma 
   \qquad \vecii{\theta_+}{\theta_-} 
   \ = \ \frac{1}{l} \matrixii{\rho_+}{-\rho_-}{-\rho_-}{\rho_+} 
       \vecii{\tau/l}{\varphi} \comma \label{tauphi}
\eqaend
and identifying $ \varphi$ to $\varphi + 2 \pi$, one obtains
the BTZ metric (\ref{BTZmetric}).

$AdS_2$ is also related to $AdS_3$. By dimensionally reducing (\ref{btz})
with respect to $\theta_L$ and making a change of variables $(r/l)^2 = z$,
 one obtains the $AdS_2$ metric (\ref{ads2z}) up to a factor 
\cite{LS,Strominger2}.
\appsubsection{dimensional reduction and scalar wave equations}
We consider a $(D+1)$-dimensional geometry of the form
\eqabegin
   ds_{D+1}^2 &=& F(r) \lbb -dt^2 + dx^2 
  + u(r) (\cosh \sigma dt - \sinh \sigma d x)^2 \rbb 
        +  g_{rr}(r) dr^2 + g_\Omega(r) d\Omega_{d+1}^2
  \comma \nn \\
   \label{D+1metric}
\eqaend
where $ D=d+3 $ and $\sigma$ corresponds to the boost parameter.
By dimensional reduction (in Einstein frame) with respect to $x$, 
one obtains the $D$-dimensional geometry
\eqabegin
   ds_D^2 &=& \lb F h_\sigma \rb^{1/(d+1)} 
   \lbb -F h_\sigma^{-1} f dt^2  + g_{rr} dr^2
    + g_\Omega d\Omega_{d+1}^2 \rbb
   \comma \label{Dmetric}
\eqaend
where $h_\sigma (r) = 1 + u(r) \sinh^2 \sigma $ and $f(r) = 1 - u(r)$.

A scalar field in $D+1$ dimensions is expanded as 
$ \Phi_{D+1} = \sum \ e^{-i\omega t + i k x} \phi_m(r) \chi_m $
where $\chi_m $ is the eigenfunction of the Laplacian on $S^{d+1}$ 
with the eigenvalue $-\Lambda = -m(m+d)$.
The momentum $k$ becomes the Kaluza-Klein charge in $D$ dimensions.
The radial equation is 
\eqabegin
   \lbb \lmb (\omega^2 -k^2) + u (\omega \sinh \sigma - k \cosh \sigma)^2 \rmb 
   + \sqrt{f/\bar{g}} \ \del_r \Bigl( F \sqrt{f \bar{g}}g^{rr} \Bigr) \del_r 
   - F f g_\Omega^{-1} \Lambda \rbb \phi_m(r) &=& 0
   \comma \nn \\
   && \label{D+1eq}
\eqaend
where $\bar{g} = g_{rr}g_\Omega$.
On the other hand, an uncharged scalar field in $D$ dimensions is expanded
as $\Phi_D = \sum \ e^{-i\omega t} \phi_m(r) \chi_m $ and the radial 
wave equation is 
\eqabegin
  \lbb \omega^2 h_\sigma 
     + \sqrt{f/\bar{g}} \del_r \Bigl( F \sqrt{f \bar{g}}g^{rr} \Bigr) 
   \del_r  - F f g_\Omega^{-1} \Lambda \rbb \phi_m(r) &=& 0
  \period \label{Deq}
\eqaend
As shown for the $D=5$ and 4 charged black holes 
\cite{MS}, (\ref{D+1eq}) takes the same form as (\ref{Deq}) in terms 
of the new variables
\eqabegin
  && \omega'^2 = \omega^2 - k^2 \comma \qquad 
   e^{\pm \sigma'} = \ e^{\pm \sigma} \frac{\omega \mp k}{\omega'}
   \period \label{charged}
\eqaend 

Therefore, if $\Phi_{D+1}$ is independent of $x$ ($k=0$), the primed 
variables are the same as unprimed ones and hence the radial equation is 
invariant under the dimensional reduction (as it should be). 
Moreover, since 
(\ref{D+1eq}) is regarded as the equation for the charged scalars in 
$D$ dimensions, the discussion  
on the uncharged scalars in section 4 can be extended
to the charged case by appropriate changes of parameters.
In this case, the parameters of the corresponding 
BTZ black holes effectively change according to (\ref{charged}).
%
%
\def\thebibliography#1{\list
 {[\arabic{enumi}]}{\settowidth\labelwidth{[#1]}\leftmargin\labelwidth
  \advance\leftmargin\labelsep
  \usecounter{enumi}}
  \def\newblock{\hskip .11em plus .33em minus .07em}
  \sloppy\clubpenalty4000\widowpenalty4000
  \sfcode`\.=1000\relax}
 \let\endthebibliography=\endlist
\newpage
\csectionast{References}
%

%
\end{document}